%% file: control.tex
\documentclass{jfm}

\usepackage{graphicx,bm}
\usepackage{natbib}
\usepackage[usenames,dvipsnames]{xcolor}
\usepackage{amsmath}
\usepackage{epstopdf}

\input{symbollines.tex}

\ifCUPmtlplainloaded \else
  \checkfont{eurm10}
  \iffontfound
    \IfFileExists{upmath.sty}
      {\typeout{^^JFound AMS Euler Roman fonts on the system,
                   using the 'upmath' package.^^J}%
       \usepackage{upmath}}
      {\typeout{^^JFound AMS Euler Roman fonts on the system, but you
                   dont seem to have the}%
       \typeout{'upmath' package installed. JFM.cls can take advantage
                 of these fonts,^^Jif you use 'upmath' package.^^J}%
      }
  \else
  \fi
\fi

\usepackage[normalem]{ulem}
\newcommand{\rev}[1]{#1} 
\newcommand{\revv}[1]{#1} 

\graphicspath{{./Figures/}}

\title[Using feedback control to find nonlinear solutions]{
  Surfing the edge: using feedback control to find nonlinear solutions}

\author[A.P. Willis \textit{et al.}]%
{A.\ns P.\ns W\ls I\ls L\ls L\ls I\ls S$^{1}$%
  \thanks{Email address for correspondence: a.p.willis@sheffield.ac.uk},\ns
Y.\ns D\ls U\ls G\ls U\ls E\ls T$^2$, \ns
O.\ns O\ls M\ls E\ls L'\ls C\ls H\ls E\ls N\ls K\ls O$^3$\ns
\and M.\ns W\ls O\ls L\ls F\ls R\ls U\ls M\ls$^3$}

\affiliation{$^1$School of Mathematics and Statistics, University of Sheffield, S3 7RH, United Kingdom\\[\affilskip]
$^2$LIMSI-CNRS, UPR 3251, Universit\'e Paris-Saclay, F-91403, Orsay, France\\[\affilskip]
$^3$Weierstrass Institute, Mohrenstrasse 39, 10117 Berlin, Germany.
}

\date{\today}
\begin{document}

\maketitle

\begin{abstract}
Many transitional wall-bounded shear flows are characterised by the coexistence 
in state-space
of laminar and turbulent regimes. Probing the edge boundary between the two attractors has led in the last decade to the numerical discovery of new (unstable) solutions to the incompressible Navier--Stokes equations. However, the iterative bisection method used to compute edge states can become prohibitively costly for large systems. Here we suggest a simple feedback control strategy to stabilise edge states, hence accelerating their numerical identification by several orders of magnitude. The method is illustrated for several configurations of cylindrical pipe flow. Travelling waves solutions are identified as edge states, and can be isolated rapidly in only one short numerical run.  A new branch of solutions is also identified.
When the edge state is a periodic orbit or chaotic state, the feedback control does not converge precisely to solutions of the uncontrolled system, but 
nevertheless brings the dynamics very close to the original edge manifold in a single run. We discuss the opportunities offered by the speed and simplicity of this new method to probe the structure of both state space and parameter space. 


\end{abstract}


\section{Introduction}
In the recent years, there has been increasing evidence for the dynamical importance of exact unstable solutions, such as travelling waves or periodic orbits, for weakly turbulent flows \citep{kawahara2012,chandler2013invariant, cvitanovic2013recurrent, willis2016symmetry}. This is best understood so far for the case of subcritical wall-bounded shear flows,
where the base flow is linearly stable at flow rates where turbulence 
is also sustained.
 These unstable solutions are the only non-trivial solutions, and are all disconnected from the trivial laminar solution \citep{eckhardt2007}. 

Most solutions discovered so far appear in saddle-node bifurcations and hence arise in pairs. The solution on the lower energy branch often belongs to the laminar-turbulent separatrix, ``the edge'', which separates initial conditions that lead to relaminarisation from those that experience transition to turbulence \citep{itano_toh_2001, skufca_yorke_eckhardt_2006}. 
The attractor within the edge separatrix is labelled the ``edge state''.  
When a solution located on the edge has only one real unstable eigenvalue, it 
acts as an attractor within the edge.  There may be more 
than one edge state for a given system.
Upper branch solutions are thought to be either embedded in the turbulent attractor, or, together with the lower branch solution, bracket the turbulent dynamics \citep{gibson2008}. 

Knowledge of an invariant solution within the edge has become the first step of a common strategy to unfold the whole bifurcation diagram of the system \citep{kreilos2012periodic,avila2013streamwise,ritter2016emergence}, but identification of exact coherent states remains a  demanding task.  In the absence of a well-understood sequence of bifurcations from the base flow, i.e. when all classical continuation methods fail \citep{tuckerman2000bifurcation},
several strategies have been employed to find such solutions.
The size of the systems under study, usually with $10^4$ to $10^9$ degrees of freedom, requires specialised iterative strategies, which in particular avoid any operation involving the storage of full matrices.  A perhaps more challenging issue is that a good initial guess for the identification is usually unavailable. 
During the 1990s and early 2000s, the main strategy used to bypass this issue was homotopy (see e.g. \cite{kerswell2005nonlinearity}). This relies on an efficient root finder, typically a Newton--Raphson solver coupled to an arc-length continuation algorithm.  This method may track the solutions in parameter space, but does not help to identify unconnected branches.  To apply this method, a companion problem featuring a linear instability of the base flow must first be identified. Once a bifurcated solution is \revv{found}, it is continued nonlinearly to a solution of the original problem. 
Success requires very good intuition for suitable homotopies that might work,
plus involves the complexity of solving for the multiple systems.  It is also unclear how continued solutions participate in the dynamics. 

Later, bisection methods, which make use of timesteppers, began to gain popularity: the amplitude of an arbitrary perturbation to the base flow is rescaled repeatedly,  until a trajectory is found which for a sufficiently long time stays away from both the laminar and the turbulent state \citep{itano_toh_2001, skufca_yorke_eckhardt_2006, schneider_eckhardt_yorke_2007}. Such a trajectory is usually chaotic, but with some luck, if there exists a regular solution with only one unstable eigendirection, there is hope that the bisection algorithm will converge to it \citep{schneider2008pcf}. Imposing discrete symmetries to the dynamics often reduces the number of unstable eigendirections of the ECS contained in the associated subspace \citep{duguet_willis_kerswell_2008}. As a result the likelihood of identifying symmetry-invariant solutions is increased, but their relevance to the non-symmetric dynamics is uncertain. In the more general case where the edge trajectory is chaotic, recurrence analysis can sometimes identify an approach to a simple state, which can be converged using a good Newton solver, possibly enhanced by some globalisation method \citep{viswanath2007recurrent, duguet2008relative}. The main drawbacks of bisection methods are the hazardous chances of success, the difficulty to converge interesting recurrent parts of the dynamics using Newton methods, and more importantly their cost. Indeed one bisection requires $2^n$ individual runs until machine precision $\varepsilon_M$ is reached at $n^{th}$ iteration, i.e. $n \sim -\log_2(\varepsilon_M)$, at which point the process needs to be restarted $M$ times while no simple edge state has been reached. The total cost is hence of $Mn\ge$ 100 runs at least, with the runs getting longer as the accuracy improves. For instance, in \cite{khapko2016edge}, each bisection required a total of $Mn \approx $ 400 runs, which corresponds to  $O(10^6-10^7)$ CPU hours. This high cost makes parametric studies infeasible in practice. Other numerical methods have been suggested as alternatives to the bisection-rootfinder combination, e.g. iterative adjoint optimisation methods \citep{farazmand2016adjoint,olvera2017optimising} though they involve significant mathematical and computational complexity. 
In summary, it is always desirable to find simpler and less expensive alternative methods.


In the present paper we demonstrate how unstable edge states can be
found numerically, by introducing  into the original system a ``control'' term
that counteracts the edge instability and is able to stabilise 
unstable states without altering them significantly. This enables the system to
dynamically approach the edge state in a single simulation of the controlled system.
A key property of a control for this purpose is that it should
influence only the stability, not the structure of the {\it a priori} unknown
target, i.e.\ the control should act as non-invasively as possible. 
At the same time, it should efficiently
force a large set of initial conditions towards the states of interest. Using control
as a tool to numerically find and analyse unstable objects in a
complex dynamical system has already been successfully applied to a variety of problems. 
A classical example is the time-delayed feedback control  \citep{pyragas1992continuous}
 able to stabilise certain periodic orbits provided their period is known. Another example is Selective Frequency Damping, which, by filtering all non-zero temporal frequencies, may stabilise steady state solutions \citep{aakervik2006steady}. In the latter case, however, the steady solutions must not possess unstable non-oscillatory eigenvalues \citep{vyazmina2010PhD}, which is frequently the case for edge solutions in subcritical shear flows. Note that none of these methods is designed to target the edge manifold.


In this article we propose a remarkably simple linear feedback control able to constrain the dynamics to the edge manifold, and to stabilise invariant solutions that are stable within the edge.
This scheme has recently been applied to track unstable chimera states in systems of non-locally coupled phase oscillators \citep{SOW014,wolfrum2015regular}. In an interesting analogy to shear flows, in systems of coupled oscillators ``chimera'' are metastable chaotic states that coexist with a fully synchronized ("laminar") state \citep{panaggio2015chimera}. The chimera also appear in a saddle-node bifurcation, and the corresponding unstable lower branch also acts as \revv{an} edge state separating the synchronous from the chimera regime. 
\rev{In the current fluid context, we first need to restrict the dynamics to the low-energy 
levels characteristic of the edge manifold.
So far, the bisection method has revealed that the dynamics on the edge manifold
is relatively low-dimensional, especially compared to turbulent flow
at equivalent parameters values \citep{duguet_willis_kerswell_2008}.
We wish to take advantage of this property
to extract invariant solutions from the controlled flow,
in the same way that solutions have been extracted from edge bisections.
Our aim here, however, is to achieve this in a single
simulation, rather than via the more expensive bisection approach.}

In the following section we describe the feedback control, then
in \S3 we present the numerical set-up. In \S4 we apply the control scheme to several different pipe flow cases, first in a restricted domain and then in an extended domain allowing for axial localisation. Finally possible applications for the methods are discussed in the concluding \S5.

\section{Stabilisation of lower branch equilibria by feedback control}

\begin{figure}
\centering
\includegraphics[height=4.5cm]{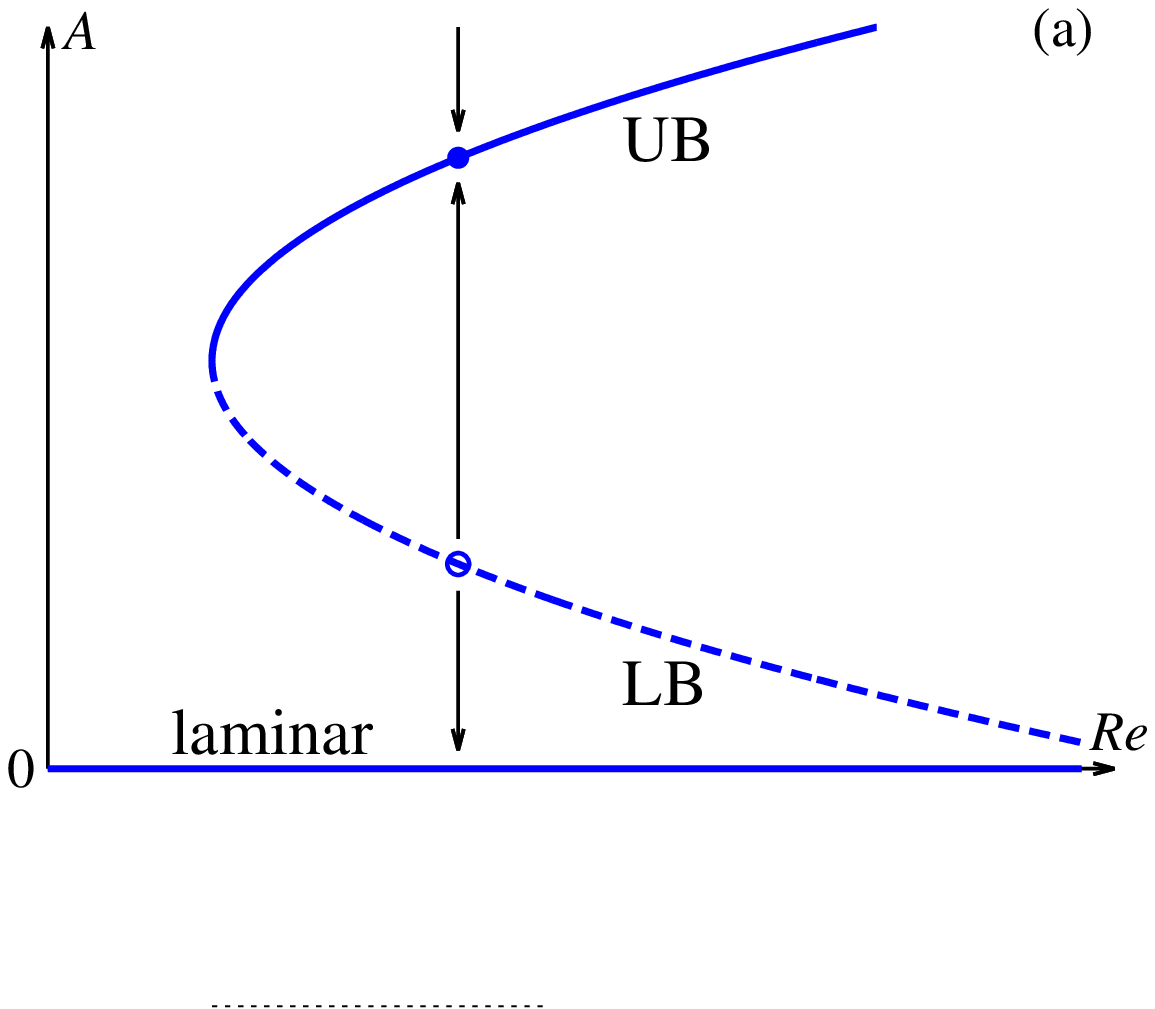}
\includegraphics[height=4.5cm]{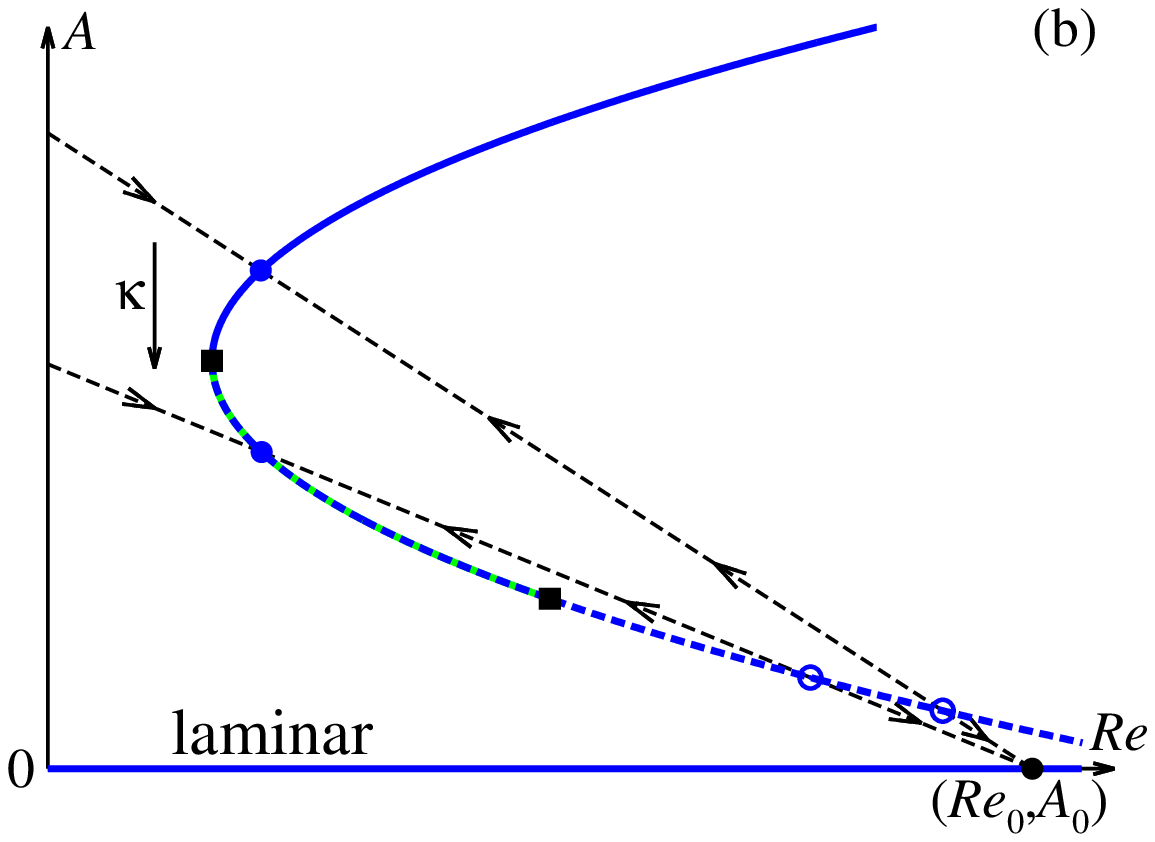}
\caption{Sketch in an $(A,Re)$ representation of the stability of the uncontrolled (a) and controlled system using the feedback control method (b) based on the constraint (\ref{constraint0}) corresponding to the dashed slanted line. The labels `UB' and `LB' refer to the stable upper and unstable lower branch in the uncontrolled system, respectively.  
In figure \ref{ft}b the dynamics is restricted to a dashed slanted line.
Rotating the line about a pivot point,
fixed points on LB are stabilised and can be tracked dynamically.}
\label{ft}
\end{figure}

In its most simple form, the control scheme makes a system parameter $\mu$
state-dependent by imposing a linear relation
between the parameter and an observable $A(t)$ 
\begin{equation}
 \mu(t) = \mu_0  + \kappa(A_0 - A(t)) \, .
 \label{constraint0}
 \end{equation}
(\rev{For the shear flow problems}
we will identify $\mu$ with $Re$ and $A$ with a component of the perturbation energy.) 
\rev{How such a proportional control can be used to stabilize the unstable part of a folded branch can be explained by 
applying it to the 
normal form for a saddle-node bifurcation, 
\begin{equation}
 \dot{x}(t) = \mu - x^2(t).        
\label{eq:1} 
\end{equation}
where equilibria lie on the parabola $\mu=x^2$.
Inserting the constraint (\ref{constraint0}) in to (\ref{eq:1}) 
and choosing the simplest possible observable
$A\equiv  x$, we arrive at the equation
\begin{equation}
\dot{x} = f(x,\kappa,\mu_0,x_0) := \mu_0 + \kappa ( x_0 - x ) - x^2.
\label{Eq:Controlled}
\end{equation}
Geometrically, the resulting dynamics can be understood
by considering its representation in the $(\mu,x)$ plane.
For $\kappa=0$ we have the uncontrolled system (\ref{eq:1}),
where for a fixed choice of the parameter the dynamics
are constrained to a vertical line $\mu=\mu_0$.
in the $(\mu,x)$ plane.  Equilibria are located at the intersections
of the vertical line with the parabola, $x=\pm\sqrt{\mu}$.
According to equation (\ref{eq:1}), $x$ increases with time 
in the region enclosed by the parabola, for $x^2<\mu$,
while it decreases outside of this region.
Hence, we conclude that for the uncontrolled system
the equilibria on the upper branch, $x=+\sqrt{\mu}$,
 of the parabola are dynamically stable, while those on the lower branch,
$x=-\sqrt{\mu}$, are unstable. 

Restricting now the dynamics to a slanted straight line,
as imposed by (\ref{constraint0}) with $\kappa>0$, we
obtain the controlled system (\ref{Eq:Controlled}), 
which has equilibria
\begin{equation}
x_\pm = \frac{1}{2} \left( -\kappa \pm \sqrt{\kappa^2 + 4 (\mu_0 + \kappa x_0)} \right),\label{equilibria}
\end{equation}
provided that
\begin{equation}
\kappa^2 + 4 (\mu_0 + \kappa x_0) = (\kappa + 2 x_0)^2 + 4 (\mu_0 - x_0^2) > 0.
\label{Discriminant}
\end{equation}
Note that the equilibria (\ref{equilibria}) coincide with the equilibria
of the uncontrolled system (\ref{eq:1}) 
for the particular choice $\mu=\mu_0+\kappa(x_0-x_\pm)$.
The controlled system has derivative
\[
\partial_x f(x_\pm,\kappa,\mu_0,x_0) = -\kappa - 2 x_\pm
= \mp \sqrt{\kappa^2 + 4 (\mu_0 + \kappa x_0)} .
\]
Thus the upper intersection $x_+$ of the controlled system is 
stable, while $x_-$ is unstable. In particular, whilst the lower branch $x=-\sqrt{\mu}$ is 
unstable for $\kappa=0$, it is possible to choose a slanted straight line
such that both intersections $x_\pm$ meet the parabola
on the lower branch.  The upper intersection $x_+$ is 
now a stablised equilibrium on the lower branch.
Further, by varying the control gain $\kappa$,
we can now sweep the straight line about the pivot point $(\mu_0,x_0)$
to track the branch of stable equilibria 
of the controlled system.  This is possible until the straight line becomes tangential to the parabola, when the $x_\pm$ disappear via a saddle-node bifurcation for
$ 
|\kappa + 2 x_0| > 2 \sqrt{ x_0^2 - \mu_0 }.
$ 

The proportional control 
(\ref{constraint0})
can be applied in the same way to a system 
with more complicated dynamics, 
shown schematically in figure \ref{ft},
\revv{provided that it contains a saddle-node bifurcation.}
Such a folded branch of flow invariant manifolds generically develops
a single transversal direction of instability, similar to the example above.
Using an observable that varies along the fold, we can apply the control.
Whenever we obtain a controlled state with constant observable,
it will exactly coincide with a dynamical state of the uncontrolled system
with fixed parameter related to the constant observable by (\ref{constraint0}).
Only the stability properties may be affected by the control,
i.e. the control is non-invasive for such states.
}
Controlled trajectories with periodic or chaotic fluctuations in $A(t)$
may still show good quantitative and qualitative agreement
with an uncontrolled trajectory for a parameter value close
to the time-average $\langle \mu \rangle_t$. 
The control may be described as non-invasive on average \citep{SOW014}.
In this way, \rev{in the shear flow problems} it will be possible
to obtain periodic orbits to be used as initial states for a corresponding Newton solver.
\rev{Also, by sweeping the control gain dynamically
from a known stable dynamical regime of the uncontrolled system,
the controlled system may yield access to qualitatively different dynamical regimes along the lower branch, i.e.\ the edge.}

\section{Numerical methodology}

\subsection{Direct numerical simulation of pipe flow}

We consider here the case of cylindrical pipe flow with a fixed mass flux. The control parameter $\mu$ in Eq. (\ref{constraint0}) is here the Reynolds number $Re=UD/\nu$, where $U$ is the constant bulk velocity, 
$D$ the pipe diameter and $\nu$ the kinematic viscosity of the fluid. 
Using $U$ and $D$ as the scales,
the non-dimensional laminar state is characterised by an axial velocity profile 
${\bm u}_\mathrm{lam}(r)=2(1-(2r)^2){\bm e}_z$ for $0\le r\le \frac{1}{2}$, driven by 
a homogeneous pressure gradient $(\partial_z p_{lam}){\bm e}_z$.
The full velocity field ${\bm u}=(u,v,w)$, which contains the radial, azimuthal and axial components of the velocity, satisfies together with the pressure $p$ the incompressible Navier--Stokes equations at all times :\begin{eqnarray}
{\bm \nabla} \cdot {\bm u}&=&0, \label{NS0} \\
\frac{\partial {\bm u}}{\partial t} + \left( {\bm u} \cdot {\bm \nabla} \right){\bm u} &=& - {\bm \nabla}p + \frac{1}{Re}{\bm \nabla}^2{\bm u}.
\label{NS}
\end{eqnarray}
The flow satisfies the no-slip velocity conditions at the wall. 
The code ensures that this condition is satisfied to machine precision
via the influence matrix method \citep{kleiser1980treatment}. 
We advance (\ref{NS}) in time using the hybrid spectral finite-difference code {\it openpipeflow.org} \citep{openpipeflow}. 
The code employs Fourier expansions 
$\mathrm{e}^{\mathrm{i}(2\alpha k z + mm_0\theta)}$, 
where $k$ and $m$ are respectively the axial and the azimuthal wavenumbers. This imposes streamwise periodicity with a wavelength $L=\pi/\alpha$ in units $D$. The integer $m_0$ indicates the degree of rotational symmetry of the flow field in the azimuthal direction, $m_0=2$ refers to a two-fold symmetry while $m_0=1$ corresponds to the absence of any discrete rotational symmetry. In some cases we also impose mirror symmetry $Z{\bm u}={\bm u}$ about the plane $\theta=0$, where
\begin{eqnarray}
Z:(u,v,w,p)(r,\theta,z)\rightarrow (u,-v,w,p)(r,-\theta,z),
\end{eqnarray}
and the shift-and-reflect symmetry $S{\bm u}={\bm u}$, where
\begin{eqnarray}
S:(u,v,w,p)(r,\theta,z)\rightarrow (u,-v,w,p)(r,-\theta,z+L/2).
\end{eqnarray}
In subsections (3.1) and (3.2) we consider simulations with $m_0=2$ in domains with respectively $\alpha=1.25$ and $0.12566$.  
With dealiasing, variables are evaluated on grids in $r\times\theta\times z$ of respectively $64\times48\times72$ and $64\times48\times576$ points.  
In subsection (3.3), for $m_0=1$ and $\alpha=1.25$, the resolution is $64\times96\times72$.

\subsection{Implementation of the feedback control.}

\rev{For the Navier--Stokes system (\ref{NS0})--(\ref{NS}),
the feedback control (\ref{constraint0}) is applied
by controlling the Reynolds number
\begin{equation}
 Re(t) = Re_0  + \kappa(A_0 - A(t))
 \label{constraintReA}
 \end{equation}
using the scalar observable}
\begin{equation}
A(t) = \int_{r_i}^{r_w} 
 ( u^2 + v^2 ) \,
r \, \mathrm{d}r\, 
\mathrm{d}\theta\, \mathrm{d}z \, ,
\end{equation}
\rev{where $r_w$=0.5 corresponds to the wall and $r_i$ is set to 0.35.
This observable $A$, considered only over the near-wall annular region 
$r_i\le r\le r_w$, is a robust 
signature of the presence of active coherent structures contributing to the turbulence production.
Several other observables were examined, such as energies that include
all components of the velocity, integrated over the whole domain, or the energy
input related to the average wall drag.  No significant difference between those observables was noted in the application
of the method, which indicates the robustness of the control 
approach.
}

The factor $1/Re$ appears as the coefficient of the viscous term, 
which is treated implicitly in our timestepping scheme, and the factor
is used in the evaluation of 
time stepping matrices.
\rev{
Now that $Re$ depends on time,
rather than recalculating these matrices every time step, we consider the governing equation (\ref{NS}) in the form
\begin{eqnarray}
\frac{\partial {\bm u}}{\partial t} 
-\frac{1}{Re_r}{\bm \nabla}^2{\bm u} 
&~=~&
\left(\frac{1}{Re}-\frac{1}{Re_r}\right)\,
{\bm \nabla}^2{\bm u}
- \left( {\bm u} \cdot {\bm \nabla} \right){\bm u} - {\bm \nabla}p \, ,
\label{NSv2}
\end{eqnarray}
where $Re_r$ is a reference value kept within 1\% of $Re(t)$.  
The diffusion term on the left-hand side of (\ref{NSv2}) is
treated implicitly.  The small correction term with coefficient $(1/Re-1/Re_r)$,
along with the rest of the terms on the right-hand side,
 is treated explicitly.  No numerical instability has been 
observed.}

\rev{
We now explain how to choose the constants $A_0$ and $Re_0$.
$A_0$ was chosen to be zero so that $Re(t)\to Re_0$ as $A(t)\to0$, i.e.\ as the observable approaches
zero $Re(t)$ approaches a maximum value. Since at low $Re$ turbulence frequently relaminarises, the value of $Re_0$ must be chosen to be sufficiently high 
to avoid such relaminarisations. It is determined indirectly from the value of $\kappa$ necessary to 
stabilise turbulence at lower $Re$ and lower amplitudes $A$:
For a trial $Re_0$ and knowing $(Re,A)$ for a given (uncontrolled) velocity field, 
from (\ref{constraint0}) we can estimate a 
starting value for $\kappa$. 
Starting controlled simulations with $\kappa$ at two or three times this value,
in order to enforce a lower $A(t)$, we then adjusted $Re_0$ to 
avoid relaminaristions.
Thereafter, the value $Re_0=10^4$ was found to be adequate for all 
further simulations, giving the pivot point $(Re_0,A_0)=(10^4,0)$.}

\rev{
Using a large value of $Re_0$ at first sight suggests that
spatial resolution issues may arise for large $Re(t)$. In practice, no fully turbulent simulation is run at values of $Re$ as high as $Re_0$.
Large $Re(t)$ occurs only for small energies $A(t)$, for which the
spectral drop-off is more rapid. It has been verified that the drop-off in energy spectra 
is sufficient that all stabilised states considered in this paper are 
well resolved.}


 \rev{Finally, we discuss the time-dependence of the control gain $\kappa(t)$. Continuation along the lower-energy branch towards increasing $Re$
requires the slope of the line given by (\ref{eq:1}) in the $(Re,A)$ plane to decrease, i.e. $\kappa(t)$ must increase with time. Once the energy $A(t)$ has dropped below its level for the uncontrolled system, it becomes more difficult to change $\kappa$ in steps without each step causing an initial wild fluctuation in $Re(t)$, which in turn  
is likely to cause immediate relaminarisation. This issue is avoided by allowing $\kappa(t)$ to vary smoothly. 
A simple exponential form for $\kappa(t)$ proved effective
\begin{equation}
\kappa(t) = \kappa_0 \, \mathrm{e}^{t/T},
\label{kappa}
\end{equation}
where $T$ is a chosen time-scale constant.  $T$ is also the expected
time-scale for the rate of decrease in the resulting controlled $A(t)$.
The approach is quite insensitive to the precise choice of $T$
\revv{(example in \S\ref{subsec:short})}.
As far as convergence to invariant solutions is concerned, 
$T$ should be longer than the decay time of the least-damped mode for the 
stabilised solution of the controlled system.  Since this is unknown, $T$ is simply chosen to be long relative to the \rev{$O(1-10)$} time scale of the variability observed in energy and disspation time series}.



\section{Results} \label{sec:numerics}
\subsection{Short periodic pipe with $m_0$=2}
\label{subsec:short}

We present first the results with the feedback control for a short pipe of length $L=2\pi/\alpha$ with $\alpha=1.25$, $m_0=2$ and no additional symmetry. This system was  considered in \citep{kerswell2007recurrence}, and was shown to possess a lower-branch travelling wave (TW) with only one unstable real eigenvalue. It was later verified in \citep{duguet_willis_kerswell_2008} that bisection in this system could identify a travelling wave as edge state, but further, that bisection could converge to two travelling waves from different families, depending on the initial condition. We describe now how the feedback controller described in the previous section successfully identified and stabilised 
another travelling wave solution on the edge. 

\begin{figure}
\centering
(a)\hspace{-5mm}\includegraphics[height=4.5cm]{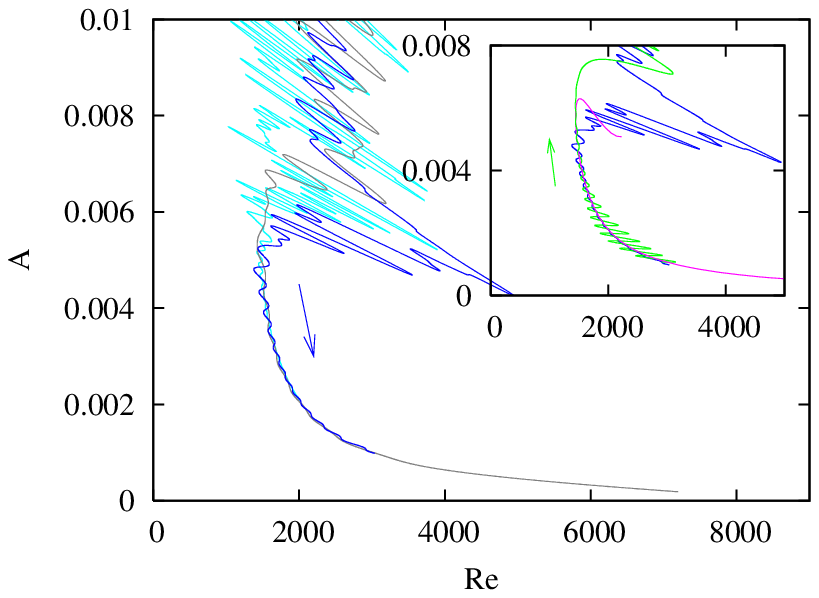}
(b)\hspace{-5mm}\includegraphics[height=4.5cm]{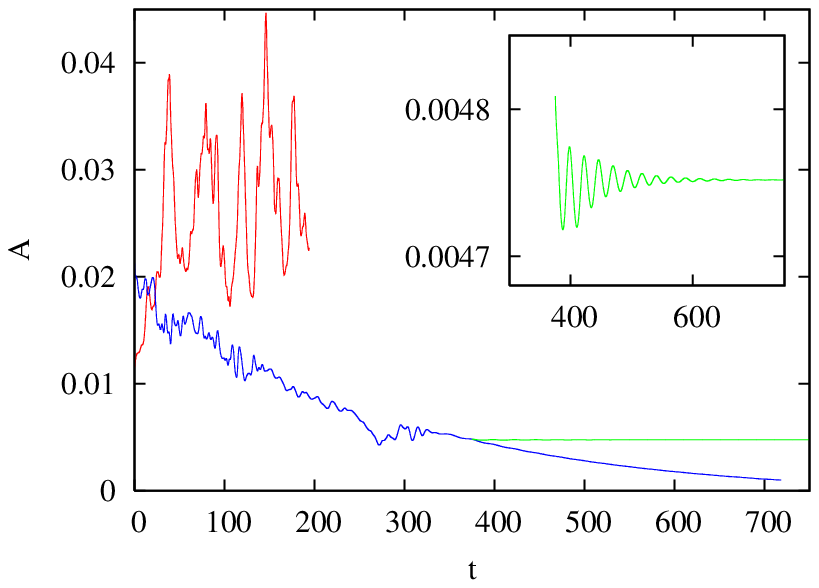}
\caption{Application of the feedback control for $\alpha=1.25$ and $m_0=2$.
(a) $(Re,A)$ representation, control with ${\kappa}=\kappa_{0}\exp{(t/T)}$,
$T=250$ (blue).  
\revv{Cases with $T$ doubled/halved converge to the same
stabilised lower branch (grey/cyan).
In the inset, the solution set is compared with results from 
arc-length continuation \revv{(using the Newton method, pink)} and 
continuation in the opposite direction using the control
with ${\kappa}=\kappa_{0}\exp{(-t/T)}$ (green).}
(b) $A(t)$ uncontrolled (red), controlled with time-varying $\kappa(t)$ (blue) and controlled with constant $\kappa$ for the stabilisation of the TW (green)
\revv{(zoom in inset).}}
\label{shortA}
\end{figure}

\begin{figure}
\centering
(a)\hspace{-0mm}\includegraphics[height=4.5cm]{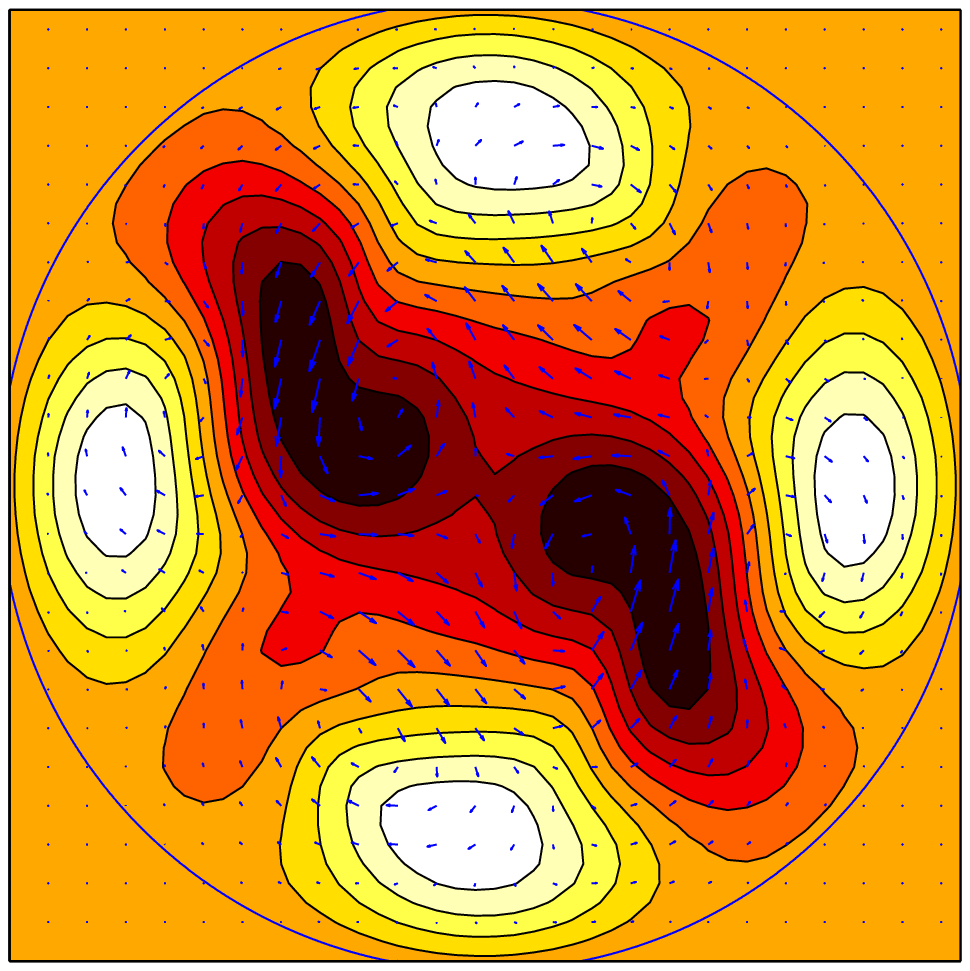}
(b)\hspace{-5mm}\includegraphics[height=4.5cm]{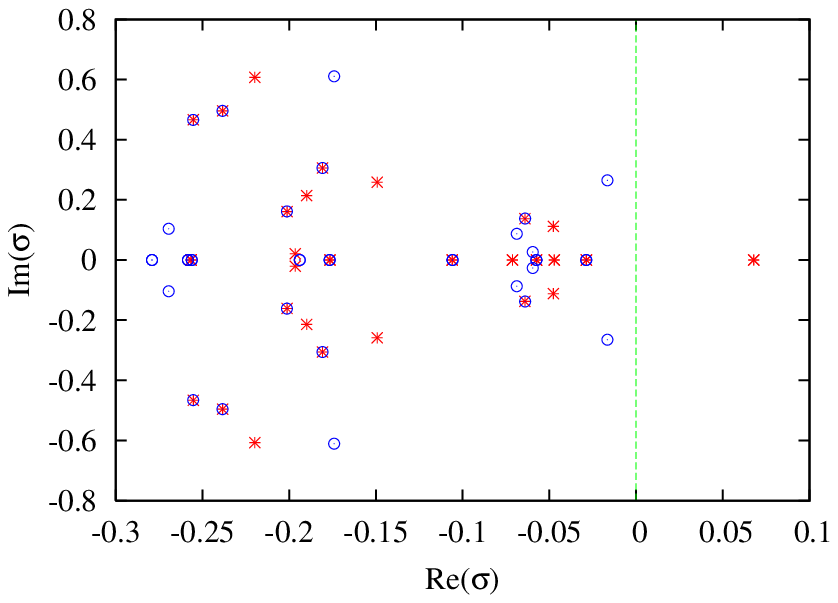}
\caption{(a) Cross-section of the stabilised TW solution with $\alpha=1.25$ and $m_0=2$. Streamwise velocity perturbation in  colour (from dark to white) and cross-stream components (vectors). (b) Eigenvalue spectrum for the TW solution with (blue circles) and without control (red crosses).}
\label{TW}
\end{figure}

We start from a turbulent snapshot at $Re=2000$,
then 
vary $\kappa(t)$ according to (\ref{kappa}) with $T=250$ and $\kappa_0=4 \times 10^5$. As shown in figures \ref{shortA}(a)--(b),
the time series of $A(t)$ initially displays erratic fluctuations until $t \approx 300~D/U$, where the dynamics appears to be smoother and fluctuations are damped. From $(A,Re) \approx (5 \times 10^{-3},1800)$ onwards, the backwards fold in figure \ref{shortA}({\it a}) indicates that the dynamics stays away from the laminar state ($A=0$) 
\revv{and that the whole lower branch is stabilised as far as the simulation
is run, up to $Re\approx3000$ (blue).
Halving $T$ (grey), more of the lower branch is quickly reached.  Convergence 
to the same branch is also observed for $T$ doubled (cyan).
For the first case, $T=250$, the state $t=375\,D/U$ is used as an initial
condition for a controlled simulation with $\dot{\kappa}$ set to zero, }
i.e.\ $\kappa$ is held fixed at $\kappa=\kappa_1=4 \times 10^5\times\mathrm{e}^{1.5}$ 
(green line in figure \ref{shortA}b). The new trajectory converges towards a fixed point, i.e.\ a travelling wave solution,
with constant $A$.
The associated velocity field in a cross-section is displayed in figure \ref{TW}(a). 
While the symmetry was not imposed, the time integration with the control converged to a solution invariant under $S$, being an S2 state of 
\citep{kerswell2007recurrence,pringle2009highly}.  It was verified using a Newton-solver \citep{willis2013revealing} that the corresponding solution is also solution to the original uncontrolled Navier--Stokes equations.  The linear stability of the states are compared in their eigenspectrum $\{\sigma\}$, calculated using an Arnoldi algorithm, with and without the control ($\kappa=\kappa_1$ and $\kappa=0$, respectively), figure \ref{TW}(b), where $Re(\sigma)>0$ indicates instability.  Whereas the TW is unstable in the uncontrolled system, it is stabilised by the feedback control. Applying the control with $\dot{\kappa}<0$ 
\rev{(here $\kappa(t)=\kappa_0\,\mathrm{e}^{-t/T}$)}
the simulation also remains on the stabilised lower branch, as demonstrated by the green line in figure \ref{shortA}(a).  \revv{Sweeping} along the lower branch in either direction is therefore a very efficient alternative to numerical continuation with a Newton scheme
for rapid numerical continuation of the TW solution in this region.

\subsection{Long periodic pipe with $m_0$=2}

\begin{figure}
\centering
(a)\hspace{-5mm}\includegraphics[height=4.5cm]{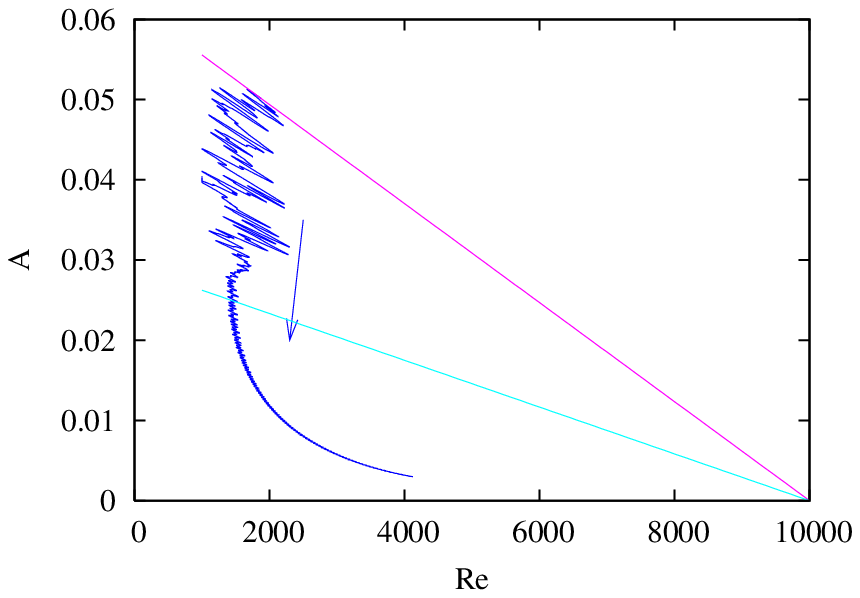}
(b)\hspace{-5mm}\includegraphics[height=4.5cm]{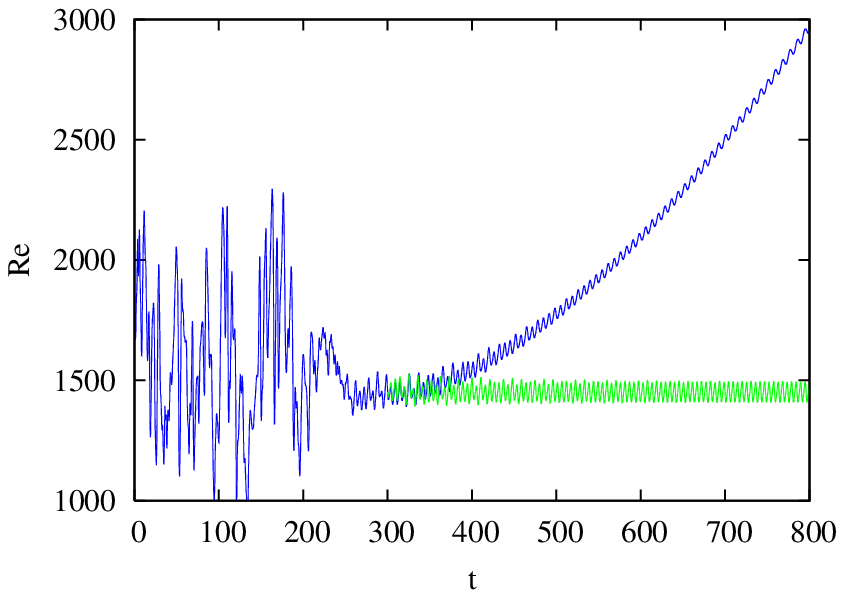}\\
(c)\hspace{-5mm}\includegraphics[height=4.5cm]{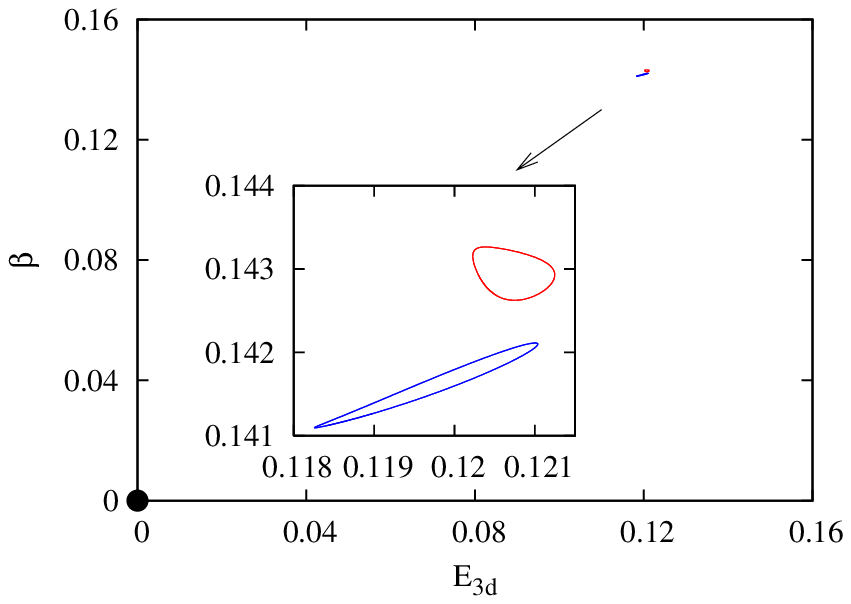}
(d)\hspace{-5mm}\includegraphics[height=4.5cm]{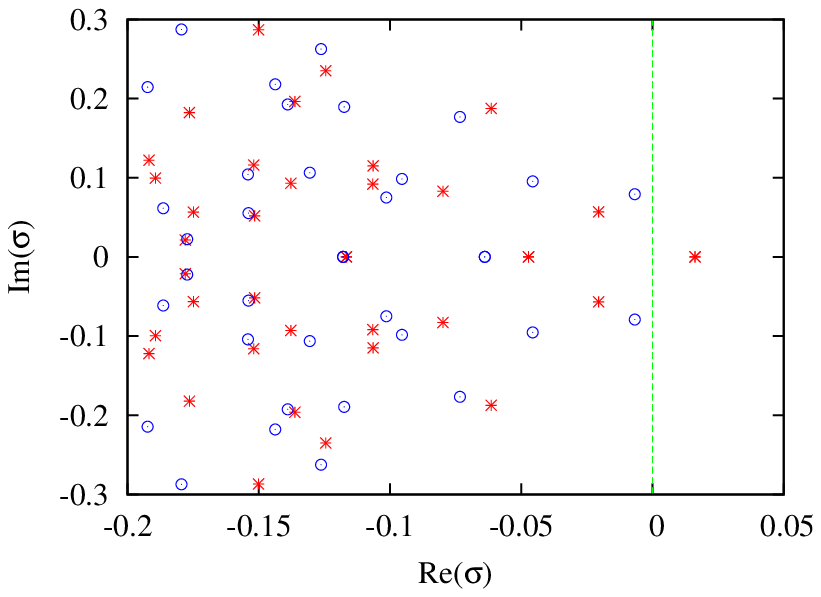}
\caption{Application of the feedback control with $L=25D$, $m_0=2$ and mirror-symmetry imposed. (a) $(Re,A)$ representation. The arrow shows the direction of rotation of the slanted straight line, displayed here are two different times (b) time series of $Re(t)$ with time-varying (blue) and constant (green) $\kappa(t)$ for the stabilisation of the RPO. 
\rev{(c) $(E_{3d},\beta)$ representation of the RPO converged using the feedback control (red), the unstable RPO of the uncontrolled system (blue) at $Re=1450$, close to each other relative to the}
laminar state (black circle). (d) Spectrum of Floquet exponents for the RPO solution with (blue circles) and without control (red crosses).}
\label{RPO}
\end{figure}

We consider next the case $L=25D$ with $m_0=2$ and mirror symmetry $Z$. This system, considered by \cite{avila2013streamwise} and later by \cite{chantry2014genesis}, is known to possess a relative periodic orbit as an edge state, in the form of a weakly modulated travelling wave whose velocity field is axially localised.  As in \S3.1, a simulation starting from a turbulent state at $Re=1900$ is first used with time-varying $\kappa(t)$ following (\ref{kappa}) with 
$\kappa_0=1.62\times10^5$ and
$T=400$. Figure \ref{RPO}(a) shows the dynamics in an $(Re,A)$ representation, where for $A \le 0.03$ a lower branch has been captured by the scheme. 
At time $t=300\,D/U$ \revv{$\kappa$ is held fixed} at $\kappa=\kappa_1=1.62\times10^5
\times\mathrm{e}^{0.75}$. 
Close-ups on the time series of $A(t)$ and $Re(t)$ show weak periodic modulations around steady values, indicating convergence towards a relative periodic orbit (RPO) (figure \ref{RPO}b). The converged RPO has a time-averaged Reynolds number $\langle Re \rangle_t \approx 1450$. 
A point on this orbit was passed to the Newton solver for the uncontrolled 
system at $Re=1450$, and converged in one Newton step.
The controlled and uncontrolled RPOs do not exactly coincide, but are close.
Figure \ref{RPO}(c) show a projection in ($E_{3d},\beta$) 
(observe the scale and distance from the origin),
where
$E_{3d}= \int |{\bm u}-\langle{\bm u}\rangle_{\theta,z}|^2\, r\,\mathrm{d}r\, \mathrm {d}\theta\, \mathrm {d}z$ and 
$1+\beta =\langle \partial_z p\rangle/\partial_z p_\mathrm{lam}$.
Their periods are respectively $10.67$ and $10.61$ units $D/U$. 
Stabilisation is substantiated in the altered Floquet exponents, displayed in figure \ref{RPO}(d). Unlike for the previous short pipe, where the edge state was a travelling wave solution, the controller has achieved convergence to an RPO, which is not precisely a solution to the uncontrolled equations. 
The control stays approximately non-invasive on average, however, and the isolated solution is close (only a single Newton step was required 
to go from one solution to the other with a relative error of less than $10^{-6}$).  
It manages this approach in only one short computation for the large 
system.

\subsection{Short periodic pipe with $m_0$=1}

\begin{figure}
\centering
(a)\hspace{-5mm}\includegraphics[height=4.5cm]{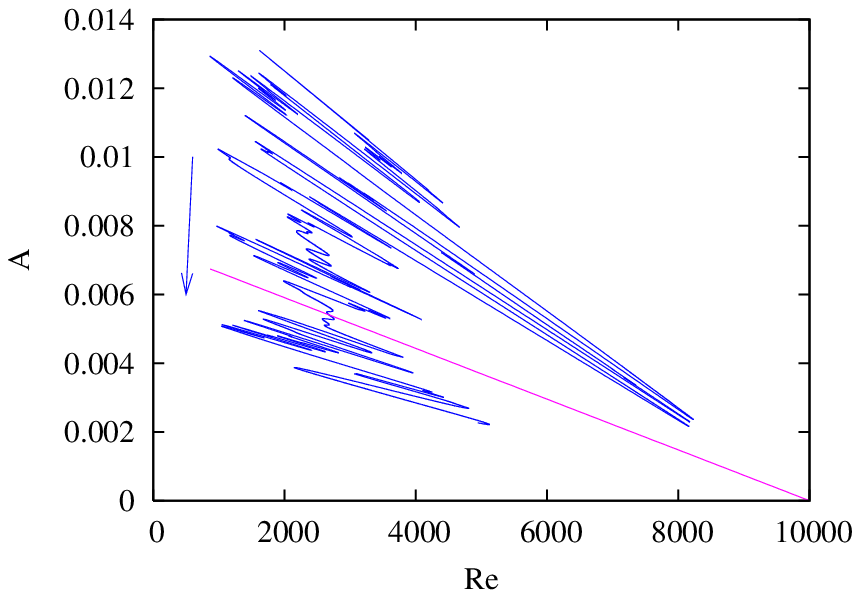}
(b)\hspace{-0mm}\includegraphics[height=4.5cm]{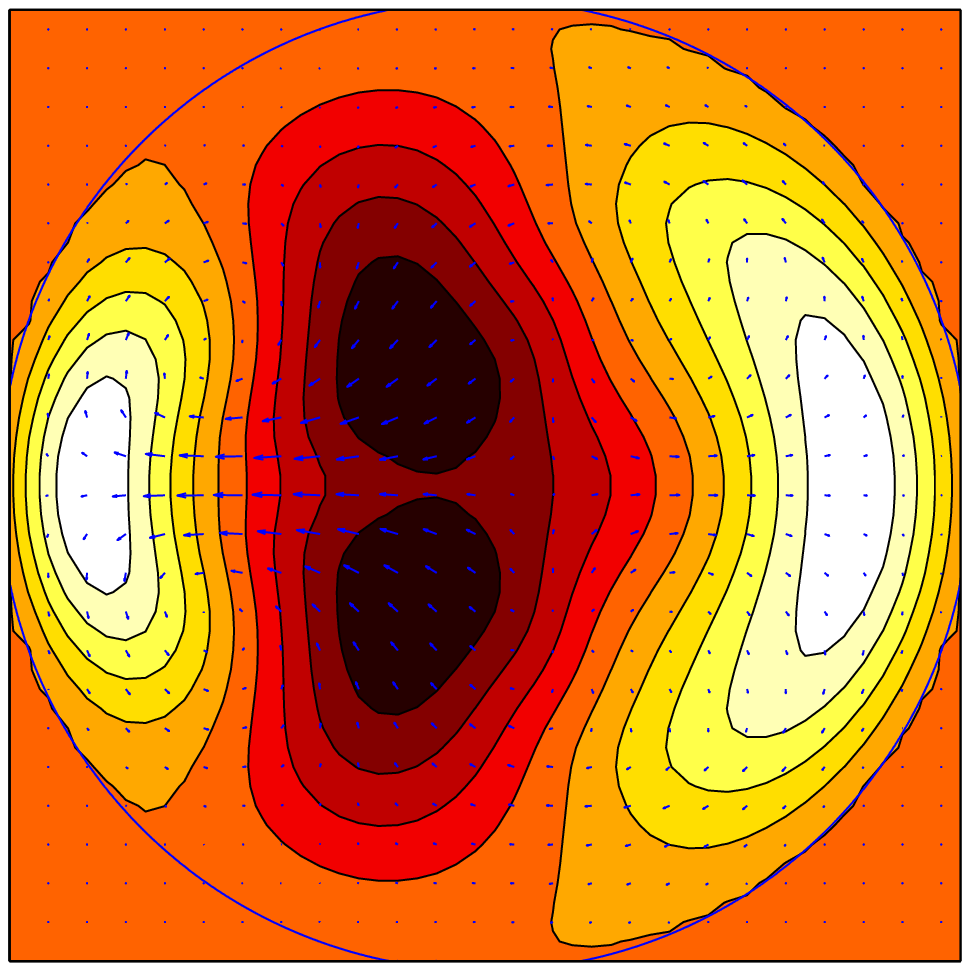}\\
(c)\hspace{-5mm}\includegraphics[height=4.5cm]{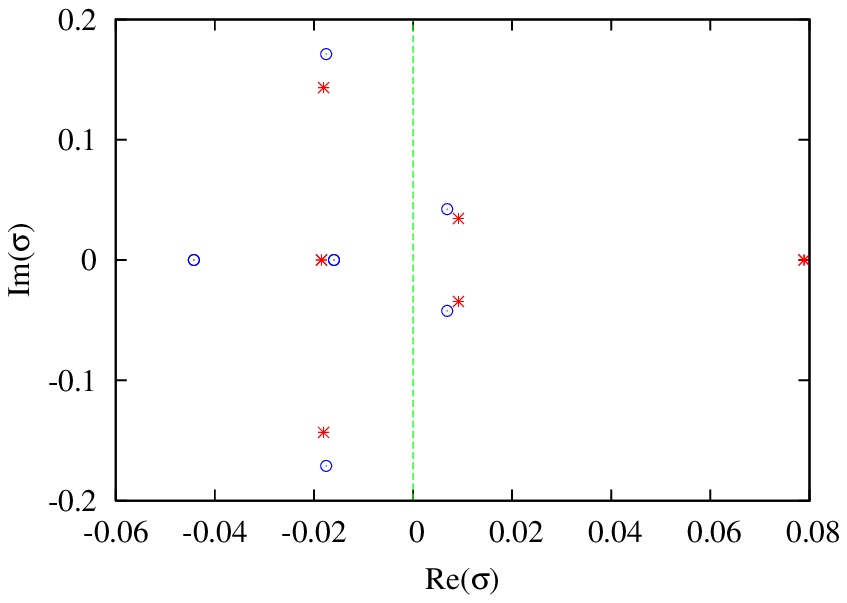}
(d)\hspace{-0mm}\includegraphics[height=4.5cm]{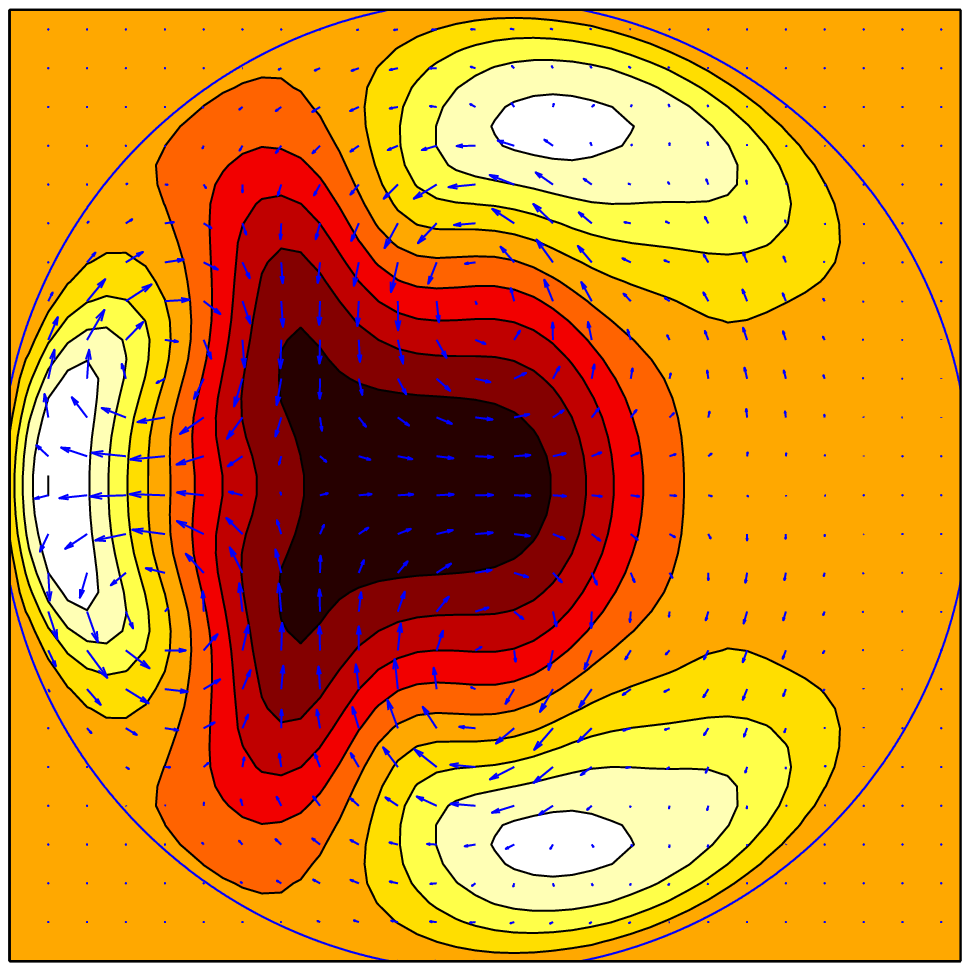}
\caption{Application of the feedback control with $\alpha=1.25$, $m_0=1$. (a) $(Re,A)$ representation.  The arrow shows the direction of rotation of the slanted straight line, which is displayed at the time where an approach to an unstable TW is identified (b) Cross-section of the stabilised TW solution (same colour coding as in figure \ref{TW}), $Re=2680$. 3. (c) Eigenvalue spectrum for the TW solution with (blue circles) and without control (red crosses).
(d) Cross-section after arc-lebgth continuation to $Re=1428$, $\alpha=1.25$.
}
\label{casem01}
\end{figure}

Bisections in periodic pipes without any rotational symmetry imposed 
($m_0=1$) have been performed 
\citep{schneider_eckhardt_yorke_2007,duguet_willis_kerswell_2008} but 
to date no edge state with a simple dynamics 
has been reported for this case 
\revv{-- the dynamics on the edge remains chaotic.}
Here we apply mirror symmetry to an $m_0=1$ computation in a short pipe with 
$\alpha=1.25$. The control by itself did not stabilise any TW solution in this case, \revv{but the dynamics on the edge is substantially less chaotic than the
turbulent state.}
A narrow window of unsteady yet quiescent dynamics of the controlled system was identified in figure \ref{casem01}(a) near the intersection with the indicated 
straight line.  Taking the state at this point as an initial guess in a Newton
search converged to a new TW solution at $Re=2680$, whose cross-section and eigenvalue spectrum are displayed in figures \ref{casem01}(b)-(c), respectively.  Note that this new TW does not satisfy the shift-and-reflect symmetry.  It is characterised by one real unstable eigenvalue and one pair of complex unstable eigenvalues. Using the control the real eigenvalue is stabilised, and the resulting solution becomes only weakly unstable. 
Also shown is a cross-section of the solution after numerical continuation to 
$Re=1428$,
\rev{where the solution appears in a saddle-node bifurcation} 
at this $\alpha=1.25$.

\section{Conclusions}

We have demonstrated that the present feedback control method is able to stabilise invariant solutions that are edge states of the uncontrolled pipe flow system. Stabilised travelling waves correspond \revv{also} to solutions of the uncontrolled Navier-Stokes equations. The control strategy is thus non-invasive in the case where the edge state of the original system is a travelling wave \citep{SOW014}. 
Relative periodic orbits or chaotic regimes stabilised using the method, however, are not precisely invariant solutions of the uncontrolled Navier--Stokes equations,
but are expected to be sufficiently close for rapid continuation \revv{towards} the 
original system.
In both cases, the feedback control method proves very efficient at bringing the system close to the original edge state in only one short computation, whereas the same task using the bisection method would require $O(100)$ times as many simulations. 

A similar feedback control has recently been considered for turbulence in stratified plane Couette flow \citep{taylor2016} to attain a target energy.  The target energy may match that of the invariant solutions, which is not known a priori in such large domains. 

Coupled with a Newton--Krylov solver, this new method, thanks to its low cost, can be used to probe the bifurcation diagram of the original system and explore the edge manifold without prior bisection. Adapting this method to other parallel shear flows, or to any other spatiotemporal system with hysteresis, is straightforward. 
Since one run is sufficient to uncover the whole lower branch parametrised by $Re$ (or potentially another parameter such as e.g.\ $\alpha$ or an additional force), it can either be used to perform wider parametric studies, or to identify the interesting regimes within the edge when multiple bifurcations occur \citep{khapko_duguet_kreilos_schlatter_eckhardt_henningson_2013}. We recommend implementation of the present control scheme as a simple precursor to the more
expensive bisection and complex Newton--Krylov implementation. 

\acknowledgements

Discussions and sections of this work were
completed at the Kavli Institute for Theoretical Physics,
supported in part under Grant No.~NSF PHY11-25915.
A.W.\ acknowledges support by EPSRC EP/P000959/1.
M.W.\ and O.O.\ acknowledge the support by DFG in the framework of SFB 910,
``Control of self-organizing nonlinear systems: theoretical methods and
concepts of application''.

\bibliographystyle{jfm}
\bibliography{paperlist}

\end{document}

%% file: symbollines.tex
\font\smallfont=cmsy10 at 10truept
\mathchardef\bigCircle="280D

\font\bigfont=cmsy10 at 14.4truept
\mathchardef\tiMes="2902        %

\font\Bigfont=cmsy10 at 17.28truept
\mathchardef\DiaMond="2A05        %
\mathchardef\cirCle="2A0E
\mathchardef\BigCircle="2A0D

\font\Bbigfont=cmsy10 at 24.88truept
\mathchardef\buLLet="2B0F

\def\bigCirc{\raise 0.3ex\hbox{$\bigCircle$}\nobreak$\,$}

\def\Bullet{\raise-0.35ex\hbox{$\buLLet$}\nobreak$\,$}

\def\triangledown{\raise 0.2em\hbox{$\bigtriangledown$}\nobreak$\,$}

\def\minisquare{\hbox{${\vcenter{
               \hrule height 0.3pt \kern-0.4pt
               \hbox{\vrule width  0.3pt height 3.0pt \kern 2.6pt
               \vrule width  0.3pt height 3.0pt} \kern-0.4pt
               \hrule height 0.3pt}}$}}
\def\ssquare{\raise 0.175ex\hbox{${\vcenter{
               \hrule height 0.5truept       \kern-0.25truept
               \hbox{\vrule width 0.5truept height 3.0truept \kern 2.75truept
                     \vrule width 0.5truept height 3.0truept} \kern-0.25truept
               \hrule height 0.5truept}}$}\nobreak$\,$}
\def\squarex{\raise 0.175ex\hbox{${\vcenter{
               \hrule height 0.8truept       \kern-1.80truept
          \hbox{\vrule width 0.8truept height 8.0truept \kern-1.95truept
                \raise 0.8truept\hbox{$\tiMes$}     \kern-6.70truept
                \vrule width 0.8truept height 8.0truept} \kern-0.80truept
               \hrule height 0.8truept}}$}\nobreak$\,$}

\def\sqbull{\raise0.175ex\hbox{\vrule height 1.4ex width 1.6ex depth 0.2ex}\nobreak$\,$}
\def\smsqbull{\raise0.175ex\hbox{\vrule height 0.8ex width 0.9ex depth 0.2ex}\nobreak$\,$}

\def\Diamondplus{${\vcenter{\vcenter{\DiaMond} \kern-10truept
                            \hbox{\vrule width .4truept}\kern -3truept
                            \hrule height .4truept}}$\nobreak$\,$}
\newcount\ndots

\def\drawline#1#2{\raise 2.5truept\vbox{\hrule width #1truept height #2truept}}
\def\moonspace#1{\hskip #1truept}

\def\Dashy{\drawline{4.00}{1.00}}     
\def\dashy{\drawline{4.00}{0.75}}     
\def\thindashy{\drawline{4.00}{0.25}}     
\def\dashyspace{\dashy\moonspace{2}}
\def\Dashyspace{\Dashy\moonspace{2}}
\def\thindashyspace{\thindashy\moonspace{2}}
\def\longdashy{\drawline{8.00}{0.75}} 
\def\thinlongdashy{\drawline{8.00}{0.25}} 
\def\longdashyspace{\longdashy\moonspace{2}}
\def\thinlongdashyspace{\thinlongdashy\moonspace{2}}

\def\dashbox{\hbox{\dashyspace}}  
\def\Dashbox{\hbox{\Dashyspace}}  
\def\dashed{\hbox {\ndots=0 \loop\ifnum\ndots<3\advance\ndots by 1
        \dashbox\repeat\dashy}\nobreak$\,$}       
\def\Dashed{\hbox {\ndots=0 \loop\ifnum\ndots<3\advance\ndots by 1
        \Dashbox\repeat\Dashy}\nobreak$\,$}       
\def\thindashbox{\hbox{\thindashyspace}}  
\def\thindashed{\hbox {\ndots=0 \loop\ifnum\ndots<3\advance\ndots by 1
        \thindashbox\repeat\thindashy}\nobreak$\,$}       
\def\thindash{\hbox {\ndots=0 \loop\ifnum\ndots<3\advance\ndots by 1
        \thindashbox\repeat\thindashy}\nobreak$\,$}       

\def\longdashbox{\hbox{\longdashyspace}}  
\def\thinlongdashbox{\hbox{\thinlongdashyspace}}  
\def\longdash{\hbox {\ndots=0 \loop\ifnum\ndots<3\advance\ndots by 1
        \longdashbox\repeat\longdashy}\nobreak$\,$}       
\def\thinlongdash{\hbox {\ndots=0 \loop\ifnum\ndots<3\advance\ndots by 1
        \thinlongdashbox\repeat\thinlongdashy}\nobreak$\,$}       

